# Automated image segmentation and division plane detection in single live *Staphylococcus aureus* cells

Adam J. M. Wollman*[1], Helen Miller*[1], Simon Foster[2], Mark C. Leake[1,3]

*Authors contributed equally

1. Biological Physical Sciences Institute (BPSI), Departments of Physics and Biology, University of York, Heslington, York, YO10 5DD.
2. Krebs Institute, Department of Molecular Biology and Biotechnology, University of Sheffield, Sheffield, S10 2TN.

3. For correspondence: mark.leake@york.ac.uk

**Abstract**

*Staphylococcus aureus* is a coccal bacterium, which divides by binary fission. After division the cells remain attached giving rise to small clusters, with a characteristic 'bunch of grapes' morphology. *S. aureus* is an important human pathogen and this, combined with the increasing prevalence of antibiotic-resistant strains, such as Methicillin Resistant *S. aureus* (MRSA), make it an excellent subject for studies of new methods of antimicrobial action. Many antibiotics, such as penicillin, prevent *S. aureus* cell division and so an understanding of this fundamental process may pave the way to the identification of novel drugs. We present here a novel image analysis framework for automated detection and segmentation of cells in *S. aureus* clusters, and identification of their cell division planes. We demonstrate the technique on GFP labelled EzrA, a protein that localises to a mid-cell plane during division and is involved in regulation of cell size and division. The algorithms may have wider applicability in detecting morphologically complex structures of fluorescently-labelled proteins within cells in other cell clusters.

**Introduction**

*Staphylococcus aureus (S. aureus)* is a bacterium that reproduces through binary fission such that the daughter cells do not fully separate from the parents and cells form into clusters. *S. aureus* is a common member of human skin microflora, especially in the nose [1,2]. However, it can cause serious infections if it reaches underlying tissues. *S. aureus* infection is a common cause of skin and lung infections and the leading cause of bacteraemia; this can be fatal, particularly if the strain is resistant to antibiotics [3]. Methicillin resistant *S. aureus* (MRSA) is resistant to beta-lactam antibiotics (e.g. penicillins and cephalosporins), now almost completely resistant to the fluoroquinolones [4,5], and is an increasing problem in hospitals, especially so for surgical procedures involving joint replacement and secondary infections arising following chemotherapy. MRSA can be treated with the glycopeptide Vancomycin, however strains have been identified with reduced susceptibility to vancomycin [6] and even complete resistance [7].

Beta-lactam antibiotics inhibit cell wall synthesis of peptidoglycan – the key protein for maintaining cell integrity against turgor pressure. They bind irreversibly to the active site of penicillin binding proteins, preventing them from building cross links in the cell wall [8,9]. Resistance to beta-lactam antibiotics evolves through altered binding sites which no longer have high affinity for the antibiotics, or via enzymatic degradation of the beta-lactam motif. A review of antibiotic resistance in *S. aureus* can be found in Chambers and Deleo [10].

New methods of arresting cell division and killing antibiotic resistant *S. aureus* must therefore be found. Cell division has been studied primarily in the rod-shaped model organisms *Bacillus subtilis (B. subtilis)* and *Escherichia coli (E. coli*), but due to its coccal shape and apparently simplified growth and division mechanisms it forms an attractive target for investigation.

Division in *S. aureus* is driven by a complex mix of several proteins, many essential, termed the divisome (discussed here [11]). The protein FtsZ forms a ring structure at a future division site at mid-cell, known as the 'Z ring'. The exact role of many of the proteins involved in division, and their essentiality in different organisms are unknown. The protein EzrA (denoted so for 'Extra Z rings A') is crucial in *S. aureus* [12,13]. In *B. subtilis* EzrA acts as an inhibitor, preventing the formation of multiple Z rings per cycle, and EzrA is also recruited to the mid-cell early in division [14]. *In vitro*, EzrA interacts with the C terminus of FtsZ which prevents it from assembling the Z ring [15,16]. The idea that an inhibitor of Z ring formation is recruited to the divisome is surprising, but in *S. aureus* EzrA was found to also regulate cell size [12,13], preventing cells from getting so large that the Z ring could not form correctly. This agrees with the finding that in *S. aureus* inhibition of division produces cells up to twice as large as normal [12,17].

The localisation of EzrA changes through the cell cycle. In *S. aureus* EzrA locates to mid-cell early in the division process [12,18]. EzrA can therefore be used to locate the division plane in the early stages of cell division. During the early stages of cell replication *S. aureus* becomes oblate rather than truly spherical. [19] To study *S. aureus* cell division, we present novel light microscopy methods and bespoke image analysis software to detect the cell division plane, cell boundaries and other morphological features.

Light microscopy has evolved into an invaluable tool for studying complex cellular processes [20] and automated analytical and computational tools aligned with other biophysical methods are invaluable in interpreting the data [21,22]. In particular, the use of fluorescence microscopy for studying complex processes has added much insight into complex molecular architectures inside living cells [23–31]. Other cell division studies in *S. aureus* have used manual segmentation and analysis [32] or relied on super-resolution images [19]. Our methods use standard epifluorescence and brightfield images, combined with image segmentation and watershedding algorithms to detect *S. aureus* cells, determine location of the cell wall, and detect cell division planes in cells containing fluorescently labelled EzrA. These techniques are also compatible with Slimfield imaging [24,29,30,33,34] which enable tracking of single-molecule complexes [35,36] and copy number quantification through deconvolution [37].

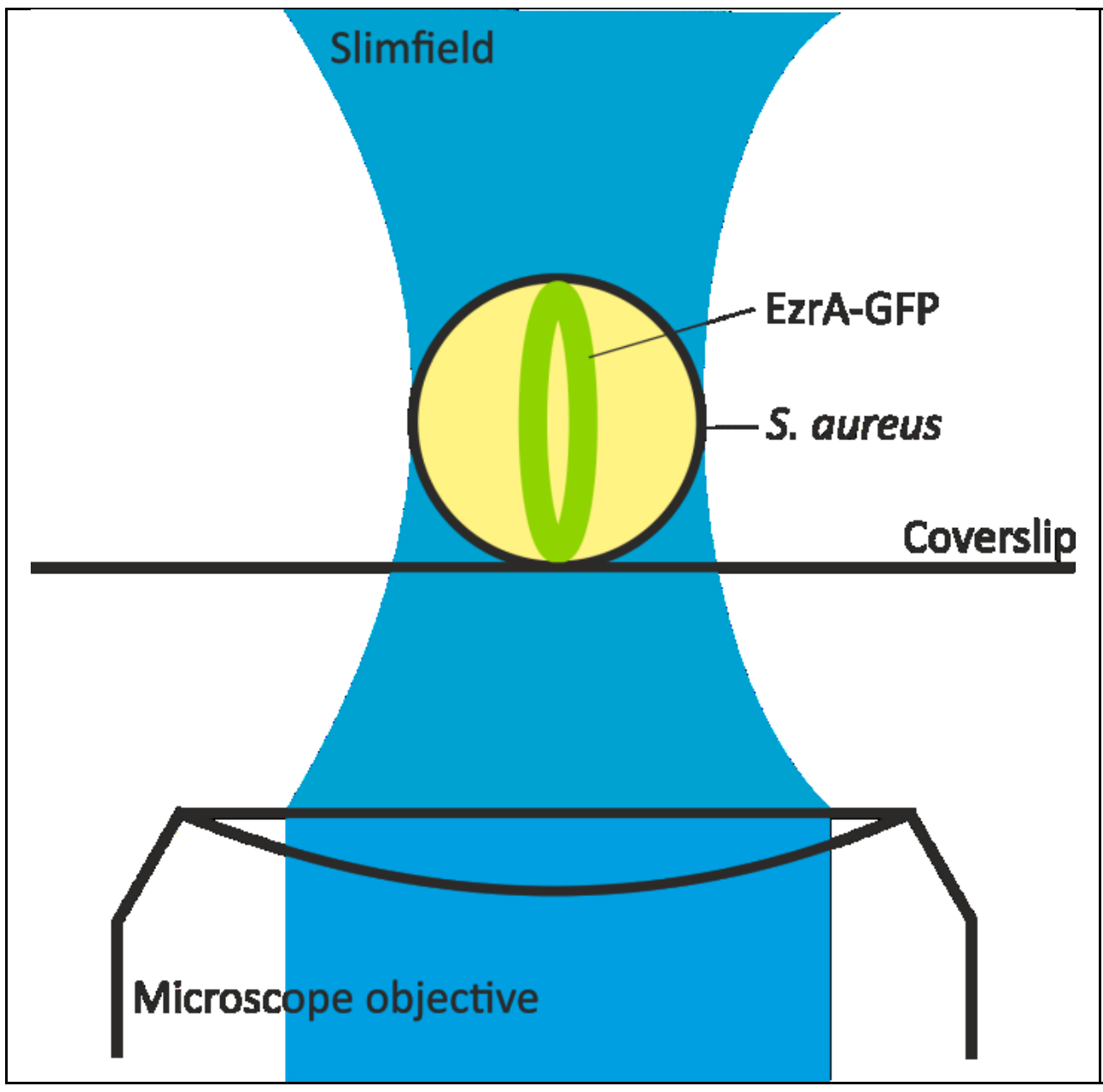

Figure 1: Schematic diagram of a *S. aureus* cell with EzrA-GFP located at mid-cell in the early stage of division, under Slimfield illumination.

**Methods**

Strains and Culturing

*S. aureus* SH1000 *ezrA-GFP*$^{+}$ (Ery$^{R}$) [12] was stored in glycerol frozen stocks at -80˚C. Cultures were grown in TSB (Tryptic soy broth) at 37˚C.

Fluorescence microscope

Our bespoke inverted fluorescence microscope was constructed from a Nikon microscope body using a 100x TIRF 1.49 NA Nikon oil immersion objective lens and a *xyz* nano positioning stage (Nanodrive, Mad City Labs). Fluorescence excitation used 50 mW Obis 488nm laser. A dual-pass GFP/mCherry dichroic with 20 nm transmission windows centred on 488 nm and 561 nm was used underneath the objective lens turret. The beam was expanded 4x, to generate an excitation field of intensity ~1.5 Wcm$^{-2}$. The beam intensity profile was measured directly by raster scanning in the focal plane while imaging a sample of fluorescent beads. A high speed camera (Photometrix Evolve Delta) was used to image at 5ms/frame with the magnification set at ~80 nm per pixel. The microscope was controlled using Micro-Manager software.

Imaging

Flow cells for imaging were constructed from standard microscope slides and plasma-cleaned coverslips by laying two lines of double-sided tape approximately 10 mm apart on the slide and dropping a coverslip onto the tape and tapping down (avoiding the imaging region), to produce a watertight linear channel [38]. The tunnel slide was coated in 0.01% poly-L-lysine to immobilise cells, and inverted for 5 minutes. This was then flushed through with 200 µl PBS buffer. Following this, a tunnel volume of cells were flushed through and the slide was left inverted for 5 minutes to allow cells to attach to the coverslip. After 5 minutes any unattached cells were washed out with 200 µl PBS buffer prior to imaging.

Computational Analysis-Segmenting cells

Brightfield and fluorescence images were segmented by defining the background intensity from the pixel intensity histogram. The density of cells in tunnel slides was chosen such that there were many more background pixels than cell pixels (fig 2a,b) and a large peak in the pixel intensity histogram at the background value. Using a threshold to find pixel values greater than the peak value plus one full width half maximum (FWHM) of the background peak finds cell-containing regions (cell clusters) very well (fig 2c,d). Morphological transformations are used to fill holes in segmented regions and remove small objects and single pixels [37].

Brightfield images alone cannot be used to find the true boundary of a cell as they are slightly defocused from fluorescence images to provide contrast, and can be misaligned from the fluorescence images. Segmenting the fluorescence image is advantageous as there is always a low-level, uniform autofluorescence in cells which gives a truer boundary between the cell and the background. This autofluorescence can be used to give the actual cell boundary [37]. The disadvantage of this method is that close-packed cells, such as the clusters typical of *S. aureus*, can be found as contiguous regions as the cells are not separated by clear regions of background intensity (fig2c). For elongated objects methods to separate overlapping cells with cell-background boundaries exist [39], but for cells with only cell-cell boundaries using the brightfield segmentations (fig2d) as seeds in a watershedding algorithm, allows the true cell boundary to be found.

Watershedding algorithms are named after river catchment basins, where ridges in the landscape form dividers (or watersheds) between catchment basins. These watersheds can be found by progressively flooding the rivers until they merge [40,41]. In the context of clustered cells, watershedding can be used to find the boundaries between neighbouring cells by flooding pixels outwards from the centres of 'seed' regions at each cell up to the boundary of the cell cluster. The fluorescence image is treated as an inverted height map such that bright pixel regions become low 'valleys'. Seeds for the algorithm, set as the lowest points in the height map are chosen, here the cell centres found from the brightfield segmentation, but automated methods exist [41]. Each seed is given a label and the pixels neighbouring the seed pixels are sorted from lowest height (pixel value) to highest. Pixels are considered in turn. If a pixel's only labelled neighbours all have the same label it is itself assigned that label, and its neighbours added to the queue at their appropriate heights. If a pixel has two neighbours with different labels, it must form part of the watershed, and is labelled as such. This continues until all pixels in the region, here the boundary of the fluorescence image segmentation, have been labelled and the watersheds are the boundaries between touching cells.

We have developed software which automatically determines the threshold of the fluorescence image and brightfield image to find masks for the outside of the cell clusters (fig. 2c)), and individual cell seeds for watershedding (fig.2d) respectively. A watershedding algorithm finds the pixels for each cell in the cluster (fig. 2e). Due to the pixelation in this step, non-physiological shapes are often recovered, so the minor and major axes, centroid position and orientation of each segmented cell pixel region are found and fitted with an ellipse (fig2f). This gives the final cell segmentation.

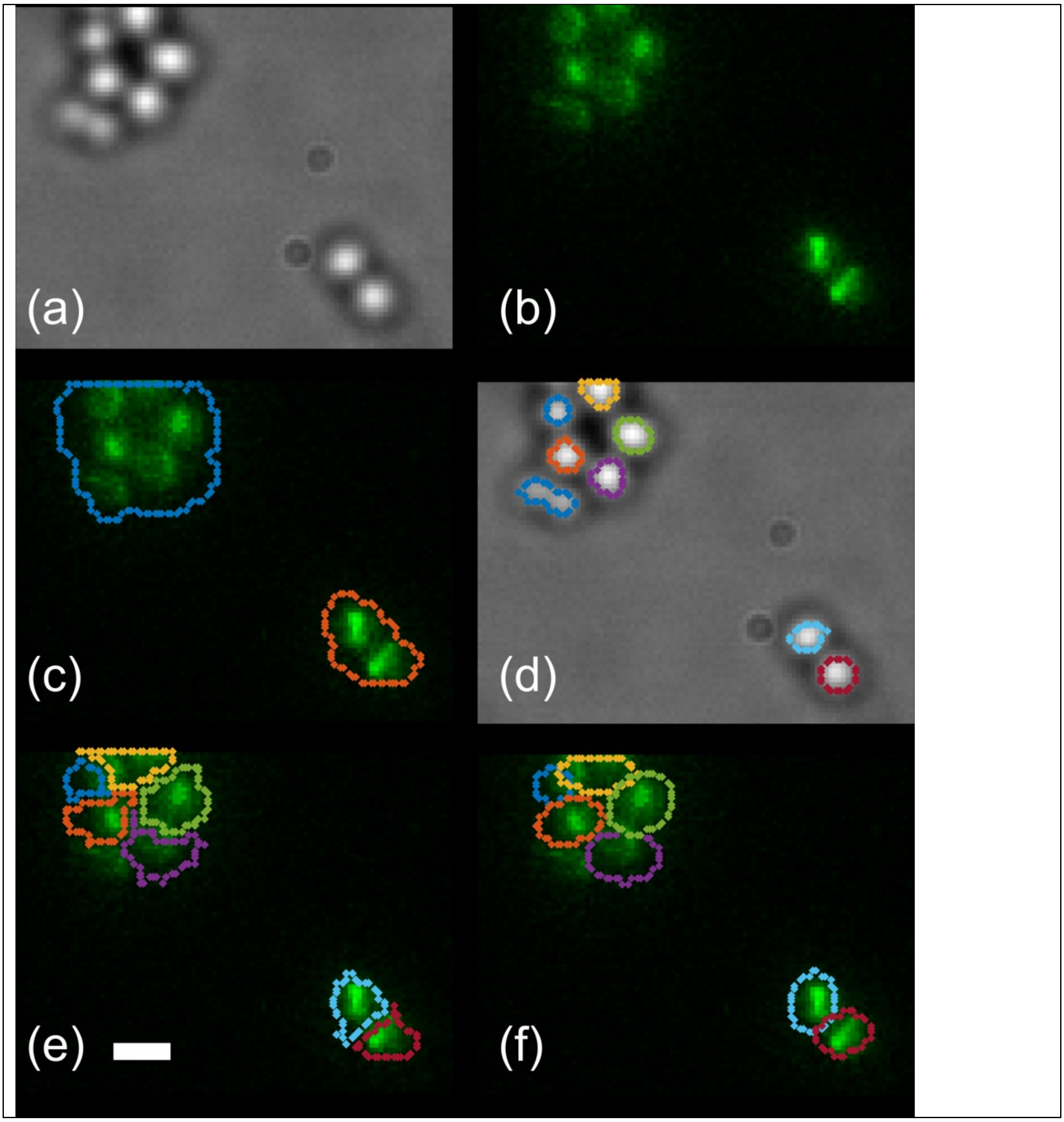

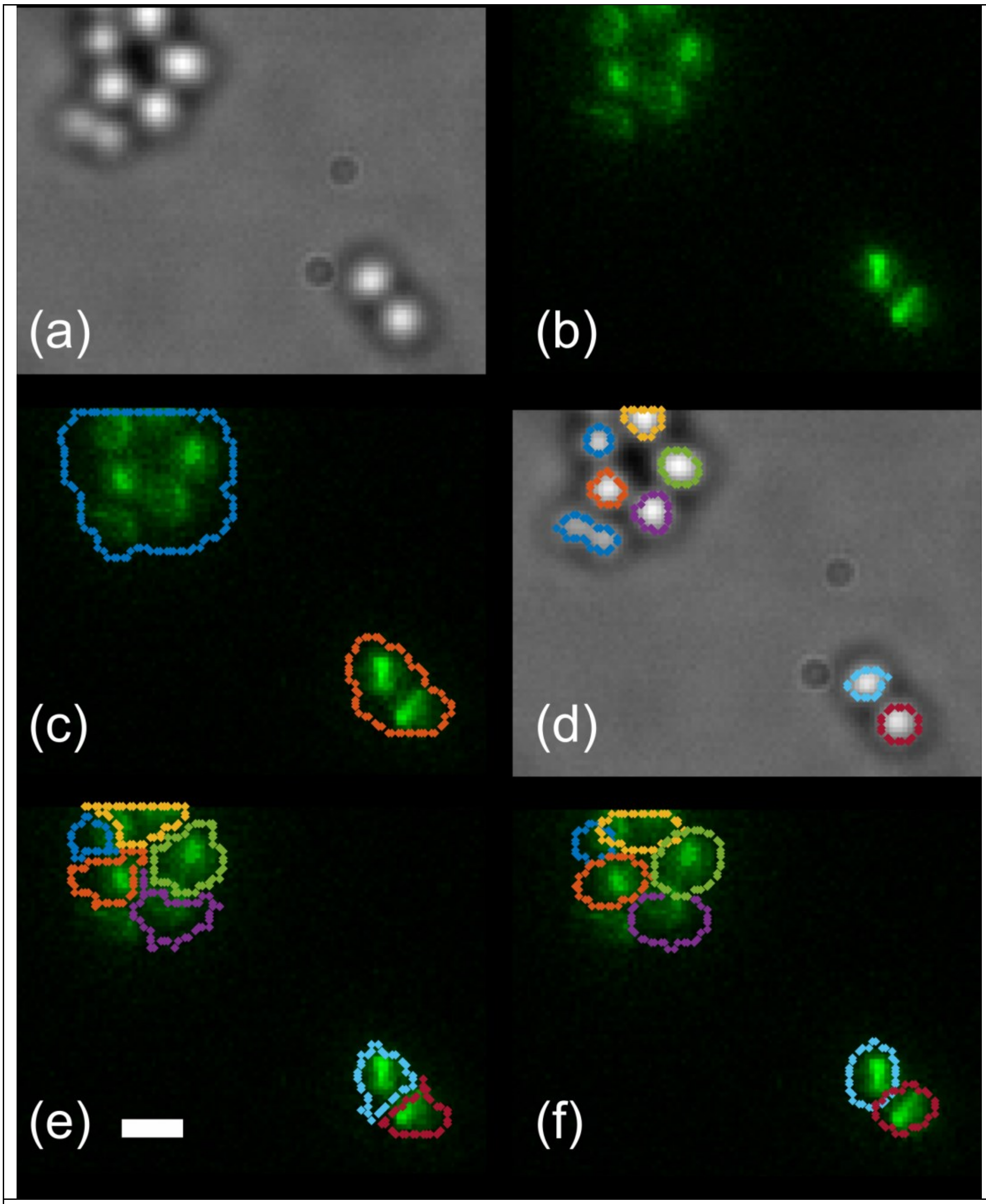

Figure 2: The cell segmentation algorithm (a) brightfield micrograph of staph cells, (b) false-colour fluorescence micrograph of EzrA-GFP, (c)segmented GFP image – cell containing regions are segmented, (d) segmented brightfield image providing individual cells and seeds for watershedding in (e), (f) watershedded cell masks used to generate ellipses.

Computational analysis- Thresholding inside cells (finding EzrA rings)

Once the pixels corresponding to cells have been identified it is relatively easy to threshold again inside cells. This is done via Otsu's method. Otsu's method [42] separates a distribution into a number of classes (here two) by minimising the intra-class variance. In the ideal case to threshold an image there are two well separated peaks, but often the valley between them is not clearly defined, due to imaging noise and difference in foreground and background pixel distributions. Otsu's method places the threshold such that the variance in each class is minimised. This method offers advantages over other methods such as fitting Gaussians [43] or valley sharpening [44] as the peaks are rarely Gaussian, and valley sharpening only considers a local area of the distribution, rather than all the data. EzrA rings appear as bright objects on the darker cell body background, and so are well suited to Otsu's method with a single threshold.

**Results**

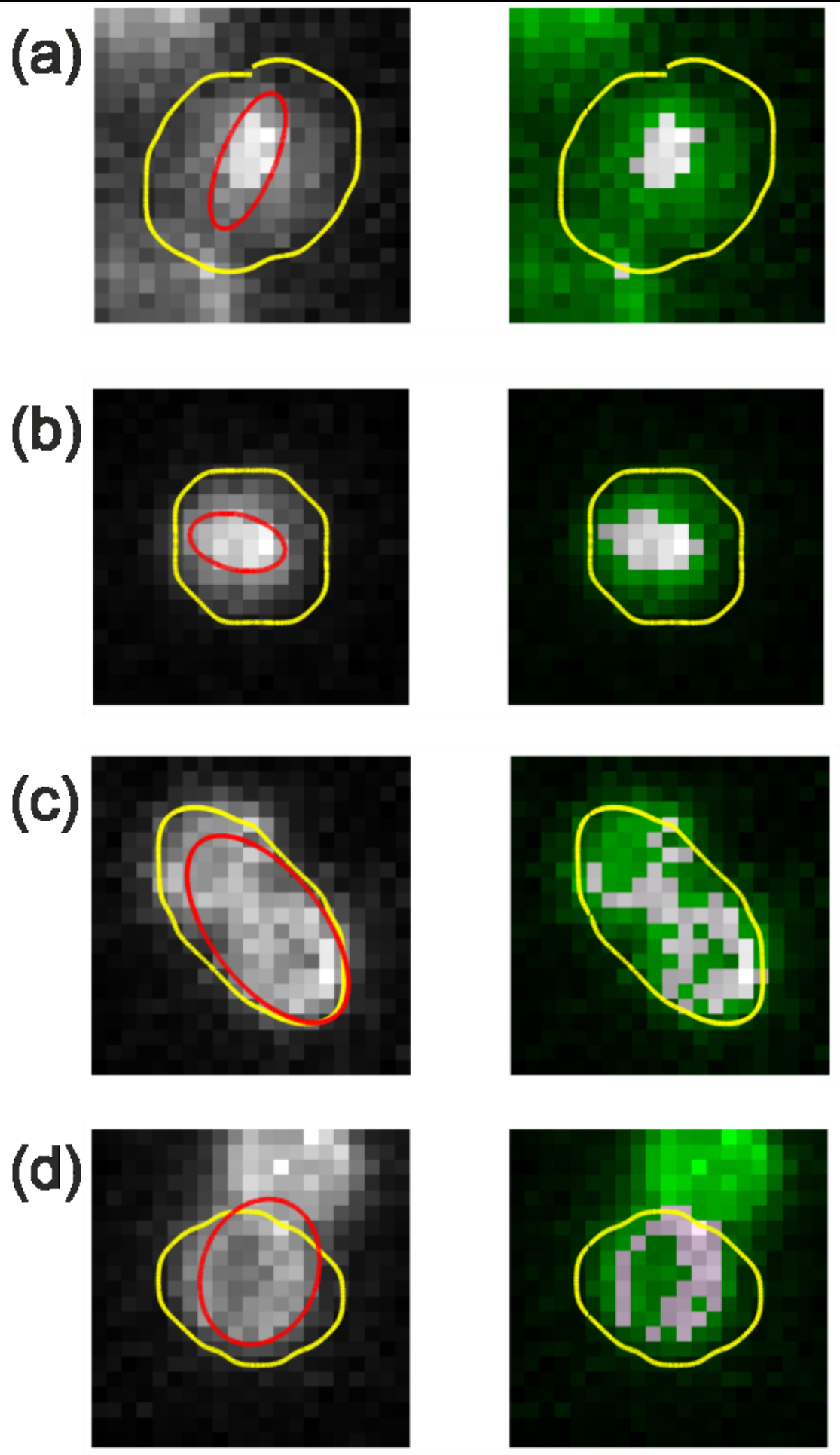

Figure 3: segmented cells, rings highlighted in white. Images on the left show the found cell envelope (yellow) and EzrA ring (red), on the right the grayscale pixels indicate the pixels EzrA was located at for the fit shown on the left. (a) & (b): examples of the algorithm detecting division planes in cells. (c) Our image segmentation algorithm has identified two divided cells at putative late stages of division and has not detected the EzrA ring. (d) shows an EzrA ring consistent with an orientation of the edge of the ring projecting towards the plane of the camera .

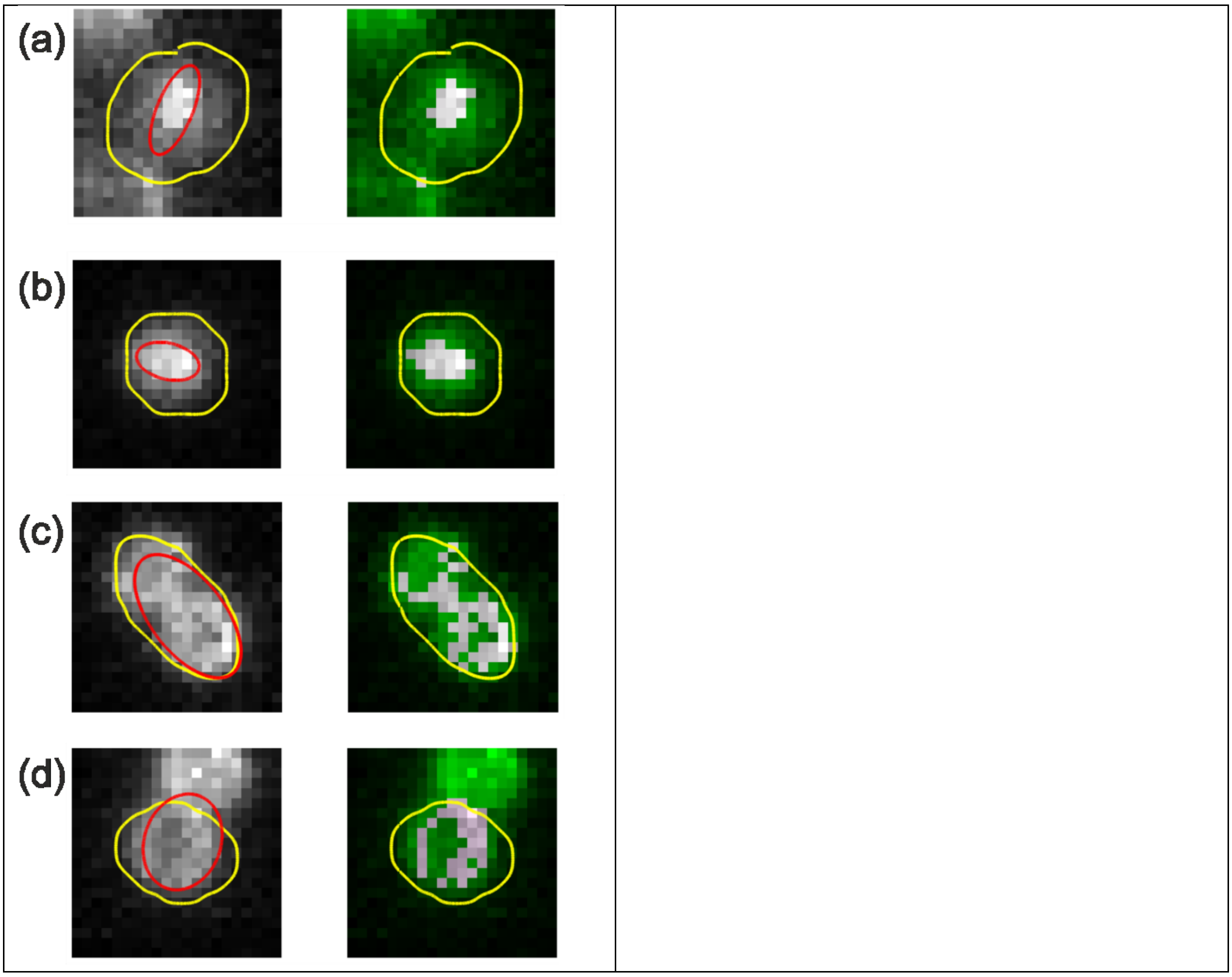

Segmenting cells

Our software successfully detected and segmented an acceptable cell mask for 34 cells out of ~60 manually counted cells. Cell masks were accepted if their pixel areas were between equivalent diameters of 0.2-2 microns. Example cell boundaries found are shown in fig. 3. Most cell boundaries are slightly elliptical (fig. 3a,b,d) with aspect ratios (ratio of major and minor axis length) approaching 1 but some extended boundaries were detected (fig. 3c) with much larger aspect ratios (fig. 4b,c). These may be pairs of cells which have been erroneously segmented together but are also likely to be dividing cells. Most of the data is close to the line of a circle with equal major and minor (fig. 4c), implying that most of the found segmentations are real elliptical cells.

The distribution of cell aspect ratios is 1.4(±0.3) (± 1s.d.) (fig 4b). A recent study by Monteiro *et al.* Reference [19] made measurements of the aspect ratio and cell dimensions using structured illumination microscopy images of vancomycin-labelled peptidoglycan in *S. aureus*. They found similar distributions of aspect ratios for cells in the P2 and P3 phases when the cells are dividing and Ezra is located at the division plane.

The distribution of cell major axis lengths is shown in fig. 4a, with mean length of 1.2(±0.3) microns, in broad agreement with other studies which do not use super-resolution methods [45,46], however, Monteiro *et al.*(2015) report cell major axis lengths at this stage in division of 0.7 microns, which is smaller than the general consensus.

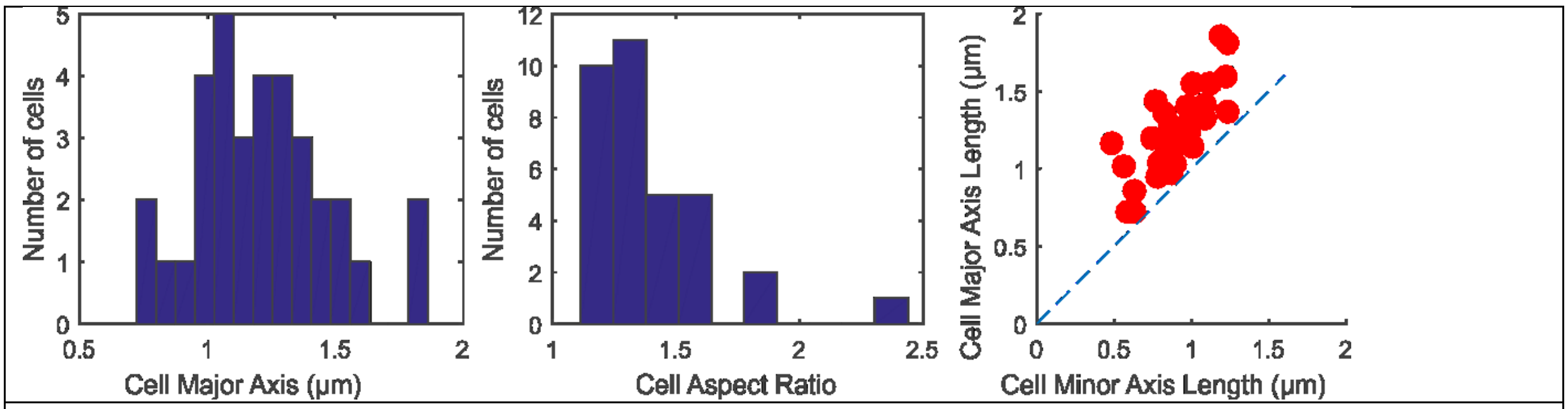

Figure 4: Distribution of cell major axis lengths, aspect ratio and scatter plot of cell major axis length against minor axis length. The dotted line lies at an aspect ratio of 1, showing many cells are elongated.

Identifying EzrA rings

A range of different shaped regions of EzrA can be found by thresholding inside the cell (fig. 3). Elliptical fits to these regions produce some thin extended ellipses but also more circular fits. The distribution of aspect ratios of these pixel regions and a scatter plot of major against minor axis length (shown in fig. 5a,b) show that ~50% of cells have extended structures with aspect ratios >>1, consistent with EzrA rings perpendicular to the image plane. The remaining structures are more circular, either corresponding to parallel rings or a completely de-localised EzrA. These can be distinguished by their areas as a function of major axis length, which accounts for projection effects. Figure 5c summarises how the area, *A*, varies as a function of major axis length, *2r,* for a fixed ring width, *w*, for a continuous circular region as produced by delocalised EzrA, a parallel ring and an ellipse produced by perpendicular EzrA ring. Most of the found areas appear consistent with rings, although a few may represent continuous structured-localised EzrA indicating that these cells are not dividing.

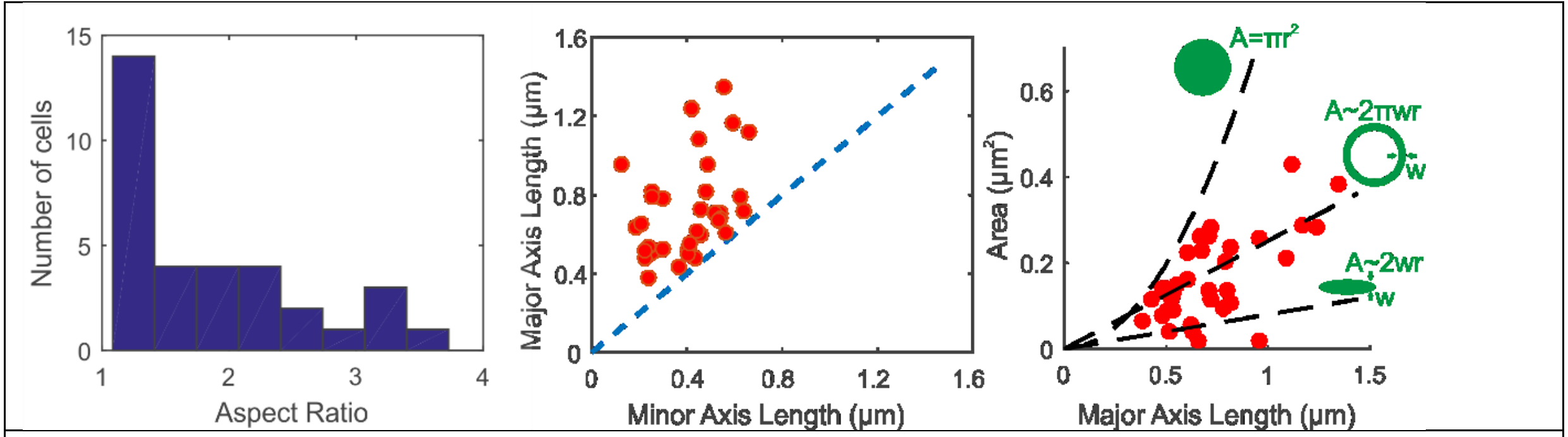

Figure 5: Left panel: Distribution of aspect ratios. Centre panel: Plot of major axis against minor axis of EzrA ring, dotted line corresponds to an aspect ratio of one for circles. Right panel: plot of area *vs* major axis length, indicates fits assuming three different models for EzrA ring morphology.

## Discussion

Our software detects cells and characterises their size and shape using an elliptical model. It then detects bright pixels inside the cell corresponding to EzrA rings, characterises their shape and determines if their orthogonality to the image plane. Our method is particularly important for *S. aureus* where cells do not move apart following division. The method can be extended to study other fluorescently labelled proteins in *S. aureus*, or in other clustering cells as it does not require the objects to be separated by background value pixels. The watershedding method using the brightfield cell centres as 'seeds' is robust to data where the brightfield image is not precisely aligned with the fluorescence image.

The aspect ratios we find for cells are in agreement with those found by Monteiro *et al.* [19], indicating that super-resolution imaging is not necessarily required to extract this parameter. Using the autofluorescence of the cell leaves another spectral channel open for protein studies. We find EzrA localised to the division plane in agreement with the expected distributions during division.

Other studies have required manual segmentation [32] or relied on super-resolution images [19] to achieve similar results. Our method is fully automated and does not require super-resolution imaging. However it is compatible with super-resolution, Slimfield microscopy and other time-resolved fluorescence localization microscopy tools which would enable tracking of single-molecule complexes [35,36,47] in EzrA rings and copy number quantification through deconvolution [37].

## Conclusion

We have written bespoke software which can segment individual *S. aureus* live cell image within cell clusters, and detect the division planes using fluorescent EzrA-GFP. It can be used to investigate cell aspect ratios, other labelled proteins that may be involved in division in *S. aureus*, and it may also have wider applicability for studying other clustering cells since it does not require cells to be separated by non-cellular background pixels. *S. aureus* is an increasing healthcare problem, particularly methicillin resistant and vancomycin resistant strains, so it is crucial to better understand its mechanism of cell division to develop new antibiotics. We hope our new image analysis tools and experimental fluorescence microscopy on live cells may aid this endeavour.

## Acknowledgements

We acknowledge technical assistance for cell preparation (Robert Turner) and image acquisition (Richard Nudd). The research was supported by the White Rose Consortium (WRC), the Biological Physical Sciences Institute (BPSI), University of York and the Medical Research Council (grant number MR/K01580X/1).

## References

[1] Kluytmans J, van Belkum A, Verbrugh H. Nasal carriage of Staphylococcus aureus: epidemiology, underlying mechanisms, and associated risks. Clin Microbiol Rev 1997;10:505–20.

[2] Gorwitz RJ, Kruszon-Moran D, McAllister SK, McQuillan G, McDougal LK, Fosheim GE, et al. Changes in the prevalence of nasal colonization with Staphylococcus aureus in the United States, 2001-2004. J Infect Dis 2008;197:1226–34. doi:10.1086/533494.

[3] Tong SYC, Davis JS, Eichenberger E, Holland TL, Fowler VG. Staphylococcus aureus infections: epidemiology, pathophysiology, clinical manifestations, and management. Clin Microbiol Rev 2015;28:603–61. doi:10.1128/CMR.00134-14.

[4] Foster J, Lentino J, Strodtman R, Divincenzo C. Comparison of in vitro activity of quinolone antibiotics and Vancomycin against gentamicin-resistant and methicillin-resistant Staphylococcus aureus by time-kill kinetic studies. Antimicrob Agents Chemother 1986;30:823–7.

[5] Miller H, Wollman AJM, Leake MC. Designing a single-molecule biophysics tool for characterizing DNA damage for techniques that kill infectious pathogens through DNA damage effects. Adv Exp Med Biol 2016;in press:In press.

[6] Hiramatsu K, Aritaka N, Hanaki H, Kawasaki S, Hosoda Y, Hori S, et al. Dissemination in Japanese hospitals of strains of Staphylococcus aureus heterogeneously resistant to vancomycin. Lancet (London, England) 1997;350:1670–3. doi:10.1016/S0140-6736(97)07324-8.

[7] Koch G, Yepes A, Förstner KU, Wermser C, Stengel ST, Modamio J, et al. Evolution of resistance to a last-resort antibiotic in Staphylococcus aureus via bacterial competition. Cell 2014;158:1060–71. doi:10.1016/j.cell.2014.06.046.

[8] Waxman DJ, Strominger JL. Penicillin-Binding Proteins and the Mechanism of Action of Beta-Lactam Antibiotics1. Annu Rev Biochem 1983;52:825–69. doi:10.1146/annurev.bi.52.070183.004141.

[9] Neu HC. The Crisis in Antibiotic Resistance. Science (80- ) 1992;257:1064–73.

[10] Chambers HF, Deleo FR. Waves of resistance: Staphylococcus aureus in the antibiotic era. Nat Rev Microbiol 2009;7:629–41. doi:10.1038/nrmicro2200.

[11] Bottomley AL, Kabli AF, Hurd AF, Turner RD, Garcia-Lara J, Foster SJ. Staphylococcus aureus DivIB is a peptidoglycan-binding protein that is required for a morphological checkpoint in cell division. Mol Microbiol 2014. doi:10.1111/mmi.12813.

[12] Steele VR, Bottomley AL, Garcia-Lara J, Kasturiarachchi J, Foster SJ. Multiple essential roles for EzrA in cell division of Staphylococcus aureus. Mol Microbiol 2011;80:542–55. doi:10.1111/j.1365-2958.2011.07591.x.

[13] Jorge AM, Hoiczyk E, Gomes JP, Pinho MG. EzrA contributes to the regulation of cell size in Staphylococcus aureus. PLoS One 2011;6:e27542. doi:10.1371/journal.pone.0027542.

[14] Levin PA, Kurtser IG, Grossman AD. Identification and characterization of a negative regulator of FtsZ ring formation in Bacillus subtilis. Proc Natl Acad Sci 1999;96:9642–

7. doi:10.1073/pnas.96.17.9642.

[15] Haeusser DP, Schwartz RL, Smith AM, Oates ME, Levin PA. EzrA prevents aberrant cell division by modulating assembly of the cytoskeletal protein FtsZ. Mol Microbiol 2004;52:801–14. doi:10.1111/j.1365-2958.2004.04016.x.

[16] Singh JK, Makde RD, Kumar V, Panda D. A membrane protein, EzrA, regulates assembly dynamics of FtsZ by interacting with the C-terminal tail of FtsZ. Biochemistry 2007;46:11013–22. doi:10.1021/bi700710j.

[17] Pinho MG, Errington J. Dispersed mode of Staphylococcus aureus cell wall synthesis in the absence of the division machinery. Mol Microbiol 2003;50:871–81. doi:10.1046/j.1365-2958.2003.03719.x.

[18] Pereira PM, Veiga H, Jorge AM, Pinho MG. Fluorescent Reporters for Studies of Cellular Localization of Proteins in Staphylococcus aureus. Appl Environ Microbiol 2010;76:4346–53. doi:10.1128/AEM.00359-10.

[19] Monteiro JM, Fernandes PB, Vaz F, Pereira AR, Tavares AC, Ferreira MT, et al. Cell shape dynamics during the staphylococcal cell cycle. Nat Commun 2015;6:8055. doi:10.1038/ncomms9055.

[20] Wollman AJM, Nudd R, Hedlund EG, Leake MC. From Animaculum to single molecules: 300 years of the light microscope. Open Biol 2015;5:150019–150019. doi:10.1098/rsob.150019.

[21] Leake MC. Analytical tools for single-molecule fluorescence imaging in cellulo. Phys Chem Chem Phys 2014;16:12635–47. doi:10.1039/c4cp00219a.

[22] Wollman AJM, Miller H, Zhou Z, Leake MC. Probing DNA interactions with proteins using a single-molecule toolbox: inside the cell, in a test tube, and in a computer. Biochem Soc Trans 2015;43:139–45.

[23] Lenn T, Leake MC, Mullineaux CW. Are Escherichia coli OXPHOS complexes concentrated in specialized zones within the plasma membrane? Biochem Soc Trans 2008;36:1032–6. doi:10.1042/BST0361032.

[24] Plank M, Wadhams GH, Leake MC. Millisecond timescale slimfield imaging and automated quantification of single fluorescent protein molecules for use in probing complex biological processes. Integr Biol (Camb) 2009;1:602–12. doi:10.1039/b907837a.

[25] Chiu S-W, Leake MC. Functioning nanomachines seen in real-time in living bacteria using single-molecule and super-resolution fluorescence imaging. Int J Mol Sci 2011;12:2518–42. doi:10.3390/ijms12042518.

[26] Robson A, Burrage K, Leake MC. Inferring diffusion in single live cells at the single-molecule level. Philos Trans R Soc Lond B Biol Sci 2013;368:20120029. doi:10.1098/rstb.2012.0029.

[27] Bryan SJ, Burroughs NJ, Shevela D, Yu J, Rupprecht E, Liu L-N, et al. Localisation and interactions of the Vipp1 protein in cyanobacteria. Mol Microbiol 2014. doi:10.1111/mmi.12826.

[28] Llorente-Garcia I, Lenn T, Erhardt H, Harriman OL, Liu L-N, Robson A, et al. Single-molecule in vivo imaging of bacterial respiratory complexes indicates delocalized oxidative phosphorylation. Biochim Biophys Acta 2014;1837:811–24. doi:10.1016/j.bbabio.2014.01.020.

[29] Reyes-Lamothe R, Sherratt DJ, Leake MC. Stoichiometry and architecture of active DNA replication machinery in Escherichia coli. Science 2010;328:498–501. doi:10.1126/science.1185757.

[30] Badrinarayanan A, Reyes-Lamothe R, Uphoff S, Leake MC, Sherratt DJ. In vivo architecture and action of bacterial structural maintenance of chromosome proteins. Science 2012;338:528–31. doi:10.1126/science.1227126.

[31] Leake MC, Greene NP, Godun RM, Granjon T, Buchanan G, Chen S, et al. Variable stoichiometry of the TatA component of the twin-arginine protein transport system observed by in vivo single-molecule imaging. Proc Natl Acad Sci U S A 2008;105:15376–81. doi:10.1073/pnas.0806338105.

[32] Wheeler R, Mesnage S, Boneca IG, Hobbs JK, Foster SJ. Super-resolution microscopy reveals cell wall dynamics and peptidoglycan architecture in ovococcal bacteria. Mol Microbiol 2011;82:1096–109. doi:10.1111/j.1365-2958.2011.07871.x.

[33] Wollman AJM, Syeda AH, McGlynn P, Leake MC. Single-molecule observation of DNA replication repair pathways in E. coli. Adv Exp Med Biol 2016;In press:In press.

[34] Wollman AJM, Leake MC. Single molecule narrowfield microscopy of protein-DNA binding dynamics in glucose signal transduction of live yeast cells. Methods Mol Biol 2016;In Press:In Press.

[35] Wollman AJM, Miller H, Zhou Z, Leake MC. Probing DNA interactions with proteins using a single-molecule toolbox: inside the cell, in a test tube and in a computer. Biochem Soc Trans 2015;43:139–45. doi:10.1042/BST20140253.

[36] Miller H, Zhou Z, Wollman AJM, Leake MC. Superresolution imaging of single DNA molecules using stochastic photoblinking of minor groove and intercalating dyes. Methods 2015;88:81–8. doi:10.1016/j.ymeth.2015.01.010.

[37] Wollman AJM, Leake MC. Millisecond single-molecule localization microscopy combined with convolution analysis and automated image segmentation to determine protein concentrations in complexly structured, functional cells, one cell at a time. Faraday Discuss 2015. doi:10.1039/c5fd00077g.

[38] Leake MC, Chandler JH, Wadhams GH, Bai F, Berry RM, Armitage JP. Stoichiometry and turnover in single, functioning membrane protein complexes. Nature 2006;443:355–8.

[39] Xue Q, Jones NS, Leake MC. A general approach for segmenting elongated and stubby biological objects: Extending a chord length transform with the Radon transform. 2010 IEEE Int. Symp. Biomed. Imaging From Nano to Macro, IEEE; 2010, p. 161–4. doi:10.1109/ISBI.2010.5490388.

[40] Vincent L, Soille P. Watersheds in digital spaces: an efficient algorithm based on immersion simulations. IEEE Trans Pattern Anal Mach Intell 1991;13:583–98.

doi:10.1109/34.87344.

[41] Meyer F. Topographic distance and watershed lines. Signal Processing 1994;38:113–25. doi:10.1016/0165-1684(94)90060-4.

[42] Otsu N. A Threshold Selection Method from Gray-Level Histograms. IEEE Trans Syst Man Cybern 1979;9:62–6.

[43] Chow CK, Kaneko T. Automatic boundary detection of the left ventricle from cineangiograms. Comput Biomed Res 1972;5:388–410. doi:10.1016/0010-4809(72)90070-5.

[44] Weszka JS, Nagel RN, Rosenfeld A. A Threshold Selection Technique. IEEE Trans Comput 1974;C-23:1322–6. doi:10.1109/T-C.1974.223858.

[45] Touhami A, Jericho MH, Beveridge TJ. Atomic force microscopy of cell growth and division in Staphylococcus aureus. J Bacteriol 2004;186:3286–95. doi:10.1128/JB.186.11.3286-3295.2004.

[46] Bailey RG, Turner RD, Mullin N, Clarke N, Foster SJ, Hobbs JK. The interplay between cell wall mechanical properties and the cell cycle in Staphylococcus aureus. Biophys J 2014;107:2538–45. doi:10.1016/j.bpj.2014.10.036.

[47] Xue Q, Leake MC. A novel multiple particle tracking algorithm for noisy in vivo data by minimal path optimization within the spatio-temporal volume. 2009 IEEE Int. Symp. Biomed. Imaging From Nano to Macro, IEEE; 2009, p. 1158–61. doi:10.1109/ISBI.2009.5193263.